\documentclass{article}
\usepackage[pdftex]{graphicx}
\begin{document}

\title{
Decoherence as an inherent characteristic of quantum mechanics
}
\author{Riuji Mochizuki\thanks{E-mail:rjmochi@tdc.ac.jp}\\
Laboratory of Physics, Tokyo Dental College,\\ 2-9-7 Kandasurugadai, Chiyoda-ku, Tokyo 101-0062, Japan }

\maketitle

\begin{abstract}
We show that it is possible to explain the quantum measurement process within the framework of quantum mechanics without any additional postulates.   The key concept of the theory is decoherence, which appears as an inherent characteristic of quantum mechanics and results from the uncertainty relation.  In contrast to environment-induced decoherence, this decoherence exists prior to a measurement being made.  To clarify our idea, we examine three elemental experiments: a Stern--Gerlach-like experiment, the Einstein--Podolsky--Rosen--Bohm (EPR--Bohm) experiment, and the double-slit experiment.  By considering the first experiment, we explain how the uncertainty relation between position and momentum introduces decoherence prior to measurement. Consideration of the EPR--Bohm experiment leads us to conclude that the correlation of the EPR pair is not a consequence of what is known as the collapse of the wave function.   Our theory of decoherence can also be applied to experiments with continuous eigenvalues, such as the double-slit experiment.  Consideration of the double-slit experiment leads us to understand how pure quantum mechanics describes the fact that quanta behave as interfering particles.
\end{abstract}

\section{Introduction}
\label{intro}

The measurement problem in quantum mechanics is one of the unresolved problems of modern physics or at least a subject of debate. In common quantum mechanics, the microscopic and macroscopic worlds need to be treated separately.  However, the fact is that these are continuations of each other. Therefore, it seems that the present formulation of quantum mechanics is insufficient to describe nature.  Nevertheless, for many decades it has been unclear how the measurement problem should be formulated, and many definitions of the measurement problem exist. 

However, in this study, we do not consider a deep discussion of what the measurement problem actually is.  Although it may appear naive, we define the measurement problem as follows: {\it Is it not possible to explain the quantum measurement process within the framework of pure quantum mechanics?}  In this study, {\it pure quantum mechanics} refers to quantum mechanics that includes Born's rule \cite{Born} but no other additional postulates, such as the projection postulate \cite{von}.  The answer to this question is generally assumed to be negative, and thus it is assumed to be necessary to adopt some modification of pure quantum mechanics.  The most traditional approach is to adopt the projection postulate.  By adopting this prescription, quantum states develop not only unitarily in obeying the Schr\"{o}dinger equation but also non-unitarily with the collapse of wave packets in the measurement process. In contrast, the collapse of the wave packet is not assumed in the many-world interpretation \cite{eve,DeW}.  In this interpretation, even macroscopic states maintain coherent superpositions.  Therefore, we dispose of the assumption that {\it one} outcome is obtained by one appropriate measurement process, which is usually regarded as a matter of course.

Decoherence \cite{Zeh2,Sch} may be regarded as the most successful theory to explain the quantum measurement process without any additional postulates.  Moreover, this is frequently applied to the many-world interpretation \cite{Sau}.
Decoherence was first proposed by Zeh et al. \cite{Zeh1,Kub}, and Zurek's important work \cite{Zurek1} on this is being actively studied.  In this theory, the observed decay of interference is explained as a result of interactions between the system and the environmental degrees of freedom.  For example, a state $|Z\rangle$ of the unified system consisting of an observed system and the measurement device is given by a superposition of two states $|a\rangle$ and $|b\rangle$:
\begin{equation}
|Z\rangle=\frac{1}{\sqrt{2}}\big(|a\rangle +|b\rangle\big).
\end{equation}
Then, we suppose that this state interacts with the state $|E_0\rangle$ representing the environmental degrees of freedom.  The interaction dynamics between these states is given by 
\begin{equation}
|a\rangle|E_0\rangle\ \rightarrow\ |a\rangle|E_a\rangle,
\end{equation}
\begin{equation}
|b\rangle|E_0\rangle\ \rightarrow\ |b\rangle|E_b\rangle.
\end{equation}
The density matrix $\hat\rho_Z$ following their interaction is
\begin{equation}
\hat\rho_Z=\frac{1}{2}\big(|a\rangle|E_a\rangle+|b\rangle|E_b\rangle\big)\big(\langle a|\langle E_a|+\langle b|\langle E_b|\big).
\end{equation}
Because the environmental degree of freedom is very large, the states $|E_a\rangle$ and $|E_b\rangle$ following the interaction are approximately orthogonal if $|a\rangle$ and $|b\rangle$ are distinguishable.  Therefore, the reduced density matrix $\hat\rho_{reZ}$ of the unified system of the observed system and the measurement device, which is obtained by tracing out the environmental degrees of freedom, becomes
\begin{equation}
\hat\rho_{reZ}=\frac{1}{2}\big(|a\rangle\langle a|+|b\rangle\langle b|\big),\label{eq:reZ}
\end{equation}
and we observe decay of the interference.  Moreover, the problem of a preferred basis has also been studied in this framework \cite{Zurek2}, and a solution was presented.

However, certain difficulties remain in this decoherence theory.  One is that the unitary interaction between the system and the environment never leads to the deletion of any interference terms.  The reduced density matrix (\ref{eq:reZ}) does not indicate that the state is in either of the two states $|a\rangle$ and $|b\rangle$, because reduced density matrices represent improper mixtures \cite{Esp} and only provide a probability distribution.  Therefore, the coherence has been delocalized into the larger system including the environment \cite{Lal}.  The next problem may be more severe.  Because the coherent terms vanish {\it after} the interaction between the system and the environment, obtaining an outcome for a superposition of states represents nothing other than violating the {\it eigenvalue--eigenstate link} \cite{Sch2}, which states that eigenvalues and eigenstates have a one-to-one correspondence excluding degeneracy. In some studies, it has been insisted that this condition is unnecessary \cite{Zurek3}.  Nevertheless, if we agree with this insistence, then we must accept that quantum mechanics is insufficient to describe nature.

As stated above, most studies in this field have yielded negative answers to the question of whether explaining the quantum measurement process within the framework of pure quantum mechanics is possible.  However, in this study, we demonstrate that the answer is in fact affirmative; i.e., we can explain the measurement process within the framework of pure quantum mechanics.  We propose a new decoherence theory, in which the uncertainty of microscopic objects leads to decoherence as an inherent characteristic of pure quantum mechanics.  Because this decoherence exists prior to a measurement, the eigenvalue--eigenstate link can be maintained.  Note that we do not intend to explain the nonlocality or the counterfactual non-definiteness of quantum mechanics [18-27] by means of other concepts, as such an ambitious attempt is beyond our scope.  What we do attempt to illustrate in this study is how measurement processes can be understood within the scope of pure quantum mechanics.

We examine three experiments in the remainder of this paper. First, in Section 2 we examine a Stern--Gerlach-like experiment with an electron to illustrate our idea of decoherence.  In Section 3, we apply our theory to an Einstein--Podolsky--Rosen (EPR) \cite{EPR} pair of electrons and show that the correlation between spatially separated particles is not a result of wave packet collapse or other such processes.  In Section 4, the double-slit experiment with electrons is examined to demonstrate our theory's effectiveness for cases with continuous eigenvalues. This also illuminates how pure quantum mechanics describes the fact that electrons behave as interfering particles, i.e., particles whose detection rate is consistent with interference.

\section{Stern--Gerlach-like experiment}
\label{sec:1}
\begin{figure}
\centering
\includegraphics[width=10cm]{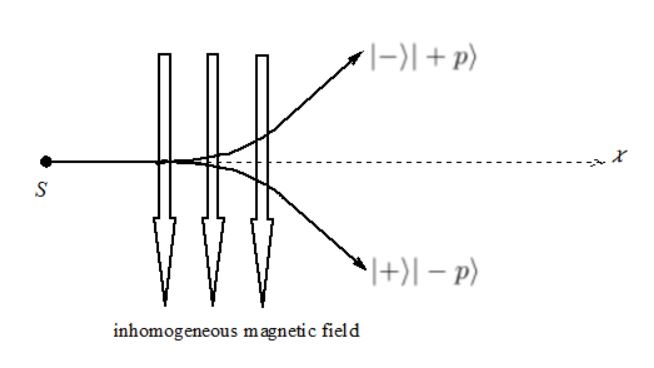}
\caption{Stern--Gerlach-like experiment}
\end{figure}
First, we examine a Stern--Gerlach-like experiment, with an electron ($S$) whose spin is measured in the $z$ direction (Fig. 1). Here, $|+\rangle$ and $|-\rangle$ are the electron's eigenstates belonging to the eigenvalues $+\hbar /2$ and $-\hbar /2$, respectively.  $S$ initially travels along the $x$ axis and enters an inhomogeneous magnetic field in the $z$ direction, where a magnetic force acts on it.  When $S$ exits the magnetic field, it has momentum $-p$ in the $z$ direction if its spin is $+\hbar /2$ and $+p$ if the spin is $-\hbar /2$.  We define the momentum eigenstates $|-p\rangle$ and $|+p\rangle$ with eigenvalues $-p$ and $+p$, respectively. Furthermore, $|0\rangle$ is defined as the momentum eigenstate with momentum 0.  Because the operators of the spin and momentum in the same direction are commutable, the state of $S$ can be described as a simultaneous eigenstate of these operators.  We define a unitary operator $\hat U_S$ that corresponds to the interaction between $S$ and the magnetic field, i.e.,  
\begin{equation}
\hat U_S|+\rangle |0\rangle =|+\rangle |-p\rangle,
\end{equation}
\begin{equation}
\hat U_S|-\rangle |0\rangle =|-\rangle |+p\rangle.
\end{equation}

 We consider the initial state $|I_S\rangle$ of $S$ defined as
\begin{equation}
|I_S\rangle\equiv\frac{1}{\sqrt{2}}\Big (|+\rangle +|-\rangle\Big )|0\rangle.
\end{equation}
Then, the state $|O_S\rangle$ after the interaction between $S$ and the magnetic field is
\begin{equation}
\begin{array}{rl}
|O_S\rangle&=\hat U_S|I_S\rangle\\
& =\frac{1}{\sqrt{2}}\Big (|+\rangle |-p\rangle +|-\rangle |+p\rangle\Big ).\label{eq:point0}
\end{array}
\end{equation}
Next, $S$ reaches one of the detectors and this records the event. If we ignore the development of $S$'s state between the magnetic field and the detector, the density matrix $\hat \rho_{S0}$ of $S$ just before detection is
\begin{equation}
\hat\rho_{S0}=|O_S\rangle\langle O_S|.\label{eq:rho0}
\end{equation}

Because we want to know $S$'s momentum in the $z$ direction, we must allow some uncertainty in its position in the same direction.  To take this uncertainty into account, we introduce the density matrix $\hat\rho_S(\zeta)$ of $S$ translated at a distance $\zeta$ in the $z$ direction, which is defined as
\begin{equation}
\hat\rho_S(\zeta)\equiv\hat T_z(\zeta)\hat\rho_{S0}\hat T_z^{\dagger}(\zeta),\label{eq:rhozeta}
\end{equation}
where $\hat T_z(\zeta)$ is the translation operator in the $z$ direction; it satisfies 
\[
\langle z+\zeta|\hat T_z(\zeta)|O_S\rangle=\langle z|O_S\rangle
\]
and is defined  by
\begin{equation}
\hat T_z(\zeta)\equiv\exp\big(-\frac{i\hat P_z\zeta}{\hbar}\Big)
\end{equation}
for the momentum operator $\hat P_z$ in the $z$ direction.

Then, we define the averaged density matrix $\hat\rho_{Sav}$ of $S$ with the uncertainty of its position in the $z$ direction as
\begin{equation}
\hat\rho_{Sav}\equiv\frac{1}{2\Delta z}\int^{\Delta z}_{-\Delta z}d\zeta\hat\rho_S(\zeta),\label{eq:integral}
\end{equation}
where, with the help of (\ref{eq:rhozeta}), we have that
\begin{equation}
\begin{array}{rl}
\hat\rho_S(\zeta)
=&\frac{1}{2}\Big (\exp{\big(\frac{+i p\zeta}{\hbar}\big)}|+\rangle |-p\rangle +\exp{\big(\frac{-ip\zeta}{\hbar}\big)}|-\rangle |+p\rangle\Big )\\
&\times \Big(\langle +p|\langle -|\exp\big(\frac{+ip\zeta}{\hbar}\big)+\langle -p|\langle +|\exp\big(\frac{-ip\zeta}{\hbar}\big)\Big).\label{eq:rho}
\end{array}
\end{equation}
Here, because we want to know what is observed with the macroscopic detector, we set
\begin{equation}
\hbar\ll p\Delta z,\label{eq:ookii}
\end{equation}
which leads to
\[
\frac{1}{2\Delta z}\int^{\Delta z}_{-\Delta z}d\zeta\exp\left(\frac{\pm 2ip\zeta}{\hbar}\right)\simeq 0.
\]
Therefore, the averaged density matrix, which describes the state of $S$ to be detected, loses its interference terms and becomes
\begin{equation}
\hat\rho_{Sav}=\frac{1}{2}\Big(|+\rangle|-p\rangle\langle -p|\langle +|+|-\rangle|+p\rangle\langle +p|\langle -|\Big).\label{eq:rhoba}
\end{equation}

If the condition (\ref{eq:ookii}) is weakened, then the interference terms in $\hat\rho_{Sav}$ would remain to a certain extent.   Performing the integral in (\ref{eq:integral}), we have that
\begin{equation}
\begin{array}{rl}
\hat\rho_{Sav}=&\frac{\hbar}{4p\Delta z}\sin \big(\frac{2p\Delta z}{\hbar}\big)\Big(|+\rangle |-p\rangle +|-\rangle |+p\rangle\Big)\Big(\langle -p|\langle +|+\langle +p|\langle -|\Big)\\
&+\frac{1}{2}\Big( 1-\frac{\hbar}{2p\Delta z}\sin\big(\frac{2p\Delta z}{\hbar}\big)\Big)\Big(|+\rangle|-p\rangle\langle -p|\langle +|+|-\rangle|+p\rangle\langle +p|\langle -|\Big).
\end{array}
\end{equation}
Because we do not adopt the projection postulate in this study, the first term in the right-hand side of this equation contributes no probability that $S$ will be found as a particle with momentum $-p$ or $+p$ in the $z$ direction.   
Therefore, there is a $1-(\hbar/2p\Delta z)\sin(2p\Delta z/\hbar)$ probability, which vanishes in the limit $\Delta z\rightarrow 0$, that $S$ will be found as a {\it particle} with momentum $-p$ or $+p$ in the $z$ direction, if we observe its position with the uncertainty $2\Delta z$. 

In studies in which it is insisted that the environment causes decoherence, the same form as (\ref{eq:rhoba}) is obtained by taking the partial trace of (\ref{eq:rho0}).  Because the state described by (\ref{eq:rho0}) is a pure state, the state described by the density matrix after taking the partial trace is an improper mixture state, which only provides the probability distribution for the outcome.  In contrast, (\ref{eq:rhoba}) represents a proper mixture state, and it describes {\it the state} itself to be detected with the macroscopic  detector.  Therefore, we can conclude that the state to be detected is not (\ref{eq:point0}) but, rather, is either $|+\rangle|-p\rangle$ or $|-\rangle|+p\rangle$.
 In this calculation, we have not used any additional postulates such as the projection postulate.  It is worth noting that this decoherence is not the result of the interaction between $S$ and the detector or other environmental factors. Rather, it is due to the uncertainty relation.

Note also that the density matrix of $S$ is not always written as (\ref{eq:rhoba}).  If we measure other observables of $S$, we must average out the density matrix over its conjugate observable.  If we observe another observable of the system, then we will obtain the averaged density matrix that has a diagonal form in this observable, as illustrated in the remainder of this section.

Suppose that an ideal macroscopic measuring device {\it M} that measures $S$'s energy is employed instead of the above-mentioned detector. Furthermore, $S$ is prepared in its neutral state $|r\rangle$ and will be at either of two energy levels $E_+$ and $E_-$, in accordance with its spin, after it exits the magnetic field. $\hat V$ is a unitary operator that changes $|r\rangle$ to $|E_+\rangle$ or $|E_-\rangle$, where these are the eigenstates whose eigenvalues are $E_+$ and $E_-$, respectively:
\begin{equation}
\hat V|+\rangle |r\rangle =|+\rangle |E_+\rangle,
\end{equation}
\begin{equation}
\hat V|-\rangle |r\rangle =|-\rangle |E_-\rangle,
\end{equation}

$|J\rangle$, which represents the initial state of their unified system, is defined as
\begin{equation}
|J\rangle\equiv\frac{1}{\sqrt{2}}\left(|+\rangle +|-\rangle\right)|r\rangle.
\end{equation}
Then, the state of $S$ just prior to being detected by $M$ is
\begin{equation}
\hat V|J\rangle =\frac{1}{\sqrt{2}}\Big (|+\rangle |E_+\rangle +|-\rangle |E_-\rangle\Big ),
\end{equation}
and its density matrix $\hat\rho_{J0}$ is given by
\begin{equation}
\hat\rho_{J0}=\hat V|J\rangle\langle J|\hat V^{\dagger}.
\end{equation}
Because we measure the energy of $S$, we require some interval of time. Therefore, we define the density matrix $\hat\rho_{Jav}$ averaged over the measurement time $2\Delta t$ as
\begin{equation}
\hat\rho_{Jav}\equiv\frac{1}{2\Delta t}\int^{\Delta t}_{-\Delta t}d\tau\exp\Big(-\frac{i\hat H\tau}{\hbar}\Big)\hat\rho_{J0}\exp\Big(+\frac{i\hat H\tau}{\hbar}\Big),
\end{equation}
where $\hat H$ is the Hamiltonian density of $S$, which satisfies
\[
\hat H|E_+\rangle =E_+|E_+\rangle,
\]
\[
\hat H|E_-\rangle =E_-|E_-\rangle.
\]
Therefore,
\begin{equation}
\hat\rho_{Jav}=\frac{1}{2\Delta t}\int^{\Delta t}_{-\Delta t}d\tau\hat\rho_J(\tau)
\end{equation}
with
\begin{equation}
\begin{array}{rl}
\hat\rho_J(\tau)=&\frac{1}{2}\big(\exp (\frac{-iE_+\tau}{\hbar})|+\rangle|E_+\rangle+\exp (\frac{-iE_-\tau}{\hbar})|-\rangle|E_-\rangle\big)\\
&\times\big(\langle E_-|\langle -|\exp (\frac{iE_-\tau}{\hbar})+\langle E_+|\langle +|\exp (\frac{iE_+\tau}{\hbar})\big).
\end{array}
\end{equation}
Here, to obtain a macroscopic result, we set
\begin{equation}
(E_+-E_-)\Delta t\gg\hbar,
\end{equation}
which leads to
\begin{equation}
\frac{1}{2\Delta t}\int^{\Delta t}_{\Delta t}d\tau\exp\Big(\frac{\pm i(E_+-E_-)\tau}{\hbar}\Big)\simeq 0.
\end{equation}
Therefore, the averaged density matrix in this case becomes
\begin{equation}
\hat\rho_{Jav}=\frac{1}{2}\Big(|+\rangle|E_+\rangle\langle E_+|\langle +|+|-\rangle|E_-\rangle\langle E_-|\langle -|\Big).\label{eq:diagonal}
\end{equation}
In this manner, (\ref{eq:diagonal}) takes a diagonal form in $S$'s energy.

\section{EPR--Bohm experiment}
\label{sec:2}
\begin{figure}
\centering
\includegraphics[width=10cm]{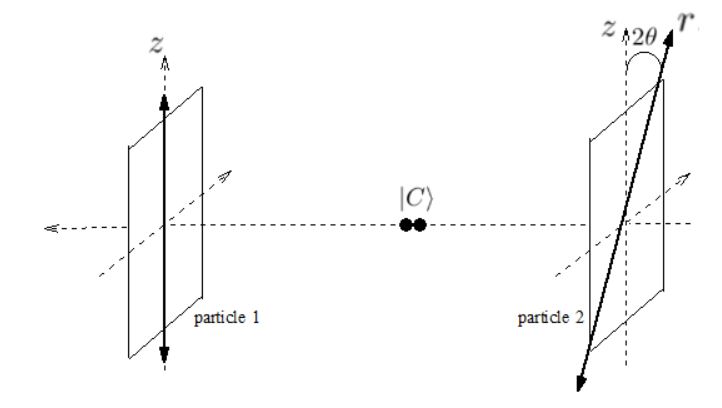}
\caption{EPR--Bohm experiment}
\end{figure}
In this section, we examine the EPR--Bohm \cite{Bohm} experiment in which two spin 1/2 particles, labeled 1 and 2, are considered (Fig. 2).  The sum of their spins should be 0, and their initial state is written as 
\begin{equation}
|C\rangle=\frac{1}{\sqrt{2}}\big(|u\rangle_1|d\rangle_2-|d\rangle_1|u\rangle_2\big),\label{eq:epr1}
\end{equation}
where $|u\rangle_i$ and $|d\rangle_i$ $(i=1,\ 2)$ are the spin eigenstates in the $z$ direction.  These satisfy
\[
(\hat\sigma_z)_i|u\rangle_i=+\frac{\hbar}{2}|u\rangle_i,
\]
\[
(\hat\sigma_z)_i|d\rangle_i=-\frac{\hbar}{2}|d\rangle_i,
\]
where $\hat\sigma_z$ is the spin operator in the $z$ direction.  The spin operator in the $r$ direction, which is perpendicular to the direction of the particles' movement and makes an angle $2\theta$ with the $z$ direction, is given by
\[
\hat\sigma_{r}=\left(
\begin{array}{cc}
\cos\theta&-\sin\theta\\
\sin\theta&\cos\theta
\end{array}
\right)
\hat\sigma_z\left(
\begin{array}{cc}
\cos\theta&\sin\theta\\
-\sin\theta&\cos\theta
\end{array}
\right).
\]
Its eigenstates $|+\rangle_i$ and $|-\rangle_i$ $(i=1,2)$ are
\begin{equation}
\begin{array}{rl}
|+\rangle_i&=\ \ \cos\theta|u\rangle_i+\sin\theta|d\rangle_i,\\
|-\rangle_i&=-\sin\theta|u\rangle_i+\cos\theta|d\rangle_i,\label{eq:epr2}
\end{array}
\end{equation}
and these satisfy
\[
(\hat\sigma_{r})_i|+\rangle_i=+\frac{\hbar}{2}|+\rangle_i,
\]
\[
(\hat\sigma_{r})_i|-\rangle_i=-\frac{\hbar}{2}|-\rangle_i
\]
By using (\ref{eq:epr1}), $|C\rangle$ can be rewritten as
\[
|C\rangle=\frac{1}{\sqrt{2}}
\big(|+\rangle_1|-\rangle_2-|-\rangle_1|+\rangle_2\big).
\]

Suppose that two observers simultaneously perform measurements of the spins of particles 1 and 2, in the directions $z$ and $r$, respectively.  Rewriting $|C\rangle$ as
\begin{equation}
\begin{array}{rl}
|C\rangle=&\frac{1}{\sqrt{2}}\big[|u\rangle_1(\sin\theta|+\rangle_2+\cos\theta|-\rangle_2)\\
&\ \ \ \ -|d\rangle_1(\cos\theta|+\rangle_2-\sin\theta|-\rangle_2)\big], 
\end{array}
\end{equation}
we easily calculate that the probabilities of observing the spins of  particles 1 and 2 as $(+\frac{\hbar}{2},\ +\frac{\hbar}{2})$, $(+\frac{\hbar}{2},\ -\frac{\hbar}{2})$, $(-\frac{\hbar}{2},\ +\frac{\hbar}{2})$, and $(-\frac{\hbar}{2},\ -\frac{\hbar}{2})$ are
\[
\langle C|(|u\rangle\langle u|)_1(|+\rangle\langle +|)_2|C\rangle=\frac{1}{2}\sin^2\theta,
\]
\[
\langle C|(|u\rangle\langle u|)_1(|-\rangle\langle -|)_2|C\rangle=\frac{1}{2}\cos^2\theta,
\]
\[
\langle C|(|d\rangle\langle d|)_1(|+\rangle\langle +|)_2|C\rangle=\frac{1}{2}\cos^2\theta,
\]
\[
\langle C|(|d\rangle\langle d|)_1(|-\rangle\langle -|)_2|C\rangle=\frac{1}{2}\sin^2\theta,
\]
respectively.

In the same manner as for the Stern--Gerlach-like experiment,  particles 1 and 2 enter inhomogeneous magnetic fields in the $z$ and $r$ directions, respectively.  When the particles exit these magnetic fields, each particle obtains a momentum in its respective direction.  Initially, neither particle should have any momentum in the $z$ or $r$ direction.  If we define their state with no momenta as $|0\rangle_i$, the state $|I_C\rangle$ before entering the magnetic fields is
\[
|I_C\rangle = |C\rangle |0\rangle_1|0\rangle_2.
\]
Then, the state $|O_C\rangle$ after the interaction between the electrons and the magnetic fields is
\begin{equation}
\begin{array}{rl}
|O_C\rangle=&\hat U_C|I_C\rangle\\
=&\frac{1}{\sqrt{2}}\big[|u\rangle_1|-p\rangle_1(\sin\theta|+\rangle_2|-q\rangle_2+\cos\theta|-\rangle_2|+q\rangle_2)\\
&\ \ -|d\rangle_1|+p\rangle_1(\cos\theta|+\rangle_2|-q\rangle_2-\sin\theta|-\rangle_2|+q\rangle_2)\big],
\end{array}
\end{equation}
where the unitary operator $\hat U_C$ corresponds to the interaction and $p$ and $q$ are the momenta in the $z$ and $r$ directions, respectively.  As discussed in the previous section, we must allow some uncertainty of the position in the corresponding direction of each particle.  Therefore, the density matrix $\hat\rho_{Cav}$ that describes the state to be measured is defined as
\begin{equation}
\hat\rho_{Cav}\equiv\frac{1}{4\Delta z\Delta r}\int^{\Delta z}_{-\Delta z}d\zeta\int^{\Delta r}_{-\Delta r}d\xi\hat\rho_C(\zeta ,\xi),\label{rhobaba}
\end{equation}
where $\hat\rho_C(\zeta ,\xi)$ is the density matrix of particle 1 translated at a distance $\zeta$ in the $z$ direction and particle 2 translated at a distance $\xi$ in the $r$ direction. By using
\[
\hat T_{1z}(\zeta)\equiv\exp\Big( -\frac{i\hat P_{1z}\zeta}{\hbar}\Big),
\]
\[
\hat T_{2r}(\xi)\equiv\exp\Big( -\frac{i\hat P_{2r}\xi}{\hbar}\Big),
\]
where $\hat P_{1z}$ and $\hat P_{2r}$ are the momentum operators in the $z$ direction of particle 1 and the $r$ direction of particle 2, respectively, $\hat\rho_C(\zeta ,\xi)$ can be written as  
\begin{equation}
\hat\rho_C(\zeta ,\xi)\equiv \hat T_{1z}(\zeta)\hat T_{2r}(\xi)|O\rangle\langle O|\hat T_{2r}^{\dagger}(\xi)\hat T_{1z}^{\dagger}(\zeta).
\end{equation}
Here, in the same manner as in the previous section, because we want to know what is observed with the macroscopic detectors, we set 
\begin{equation}
\hbar\ll p\Delta z,\ \ \hbar\ll q\Delta r,\label{eq:wookii}
\end{equation}
which leads to
\[
\frac{1}{2\Delta z}\int^{\Delta z}_{-\Delta z}d\zeta\exp\left(\frac{\pm 2ip\zeta}{\hbar}\right)\simeq 0,
\]
\[
\frac{1}{2\Delta r}\int^{\Delta r}_{-\Delta r}d\xi\exp\left(\frac{\pm 2iq\xi}{\hbar}\right)\simeq 0.
\]
Therefore, the interference terms vanish, and the averaged density matrix takes a form similar to (\ref{eq:rhoba}):
\begin{equation}
\begin{array}{rl}
\hat\rho_{Cav}=\frac{1}{2}&\big[\sin^2\theta(|u\rangle|-p\rangle\langle -p|\langle u|)_1(|+\rangle|-q\rangle\langle -q|\langle +|)_2\\
&+\cos^2\theta(|u\rangle|-p\rangle\langle -p|\langle u|)_1(|-\rangle|+q\rangle\langle +q|\langle -|)_2\\
&+\cos^2\theta(|d\rangle|+p\rangle\langle +p|\langle d|)_1(|+\rangle|-q\rangle\langle -q|\langle +|)_2\\
&+\sin^2\theta(|d\rangle|+p\rangle\langle +p|\langle d|)_1(|-\rangle|+q\rangle\langle +q|\langle -|)_2\big].\label{eq:epr3}
\end{array}
\end{equation}

Because (\ref{eq:epr3}) describes a proper mixed state, the state to be detected is not a superposition but rather {\it one} of $(|u\rangle|-p\rangle)_1(|+\rangle|-q\rangle)_2$, $(|u\rangle|-p\rangle)_1(|-\rangle|+q\rangle)_2$, $(|d\rangle|+p\rangle)_1(|+\rangle|-q\rangle)_2$, or $(|d\rangle|+p\rangle)_1(|-\rangle|+q\rangle)_2$.  It is worth noting that the correlation between the spins of the two particles should not be a result of the measurement process, such as with what is known as the collapse of the wave packet, because (\ref{eq:epr3}) 
is the density matrix of the electrons {\it prior} to the measurement process.  Therefore, we should not regard the EPR experiment as evidence of instantaneous propagation of the collapse of the wave packet. 

However, we should also not regard this as evidence for an opinion that the quantum mechanics is counterfactually definite, either.  Because we can obtain only one averaged density matrix for an observed pair of electrons, we cannot suppose the state before a measurement possesses a definite value for the spin in each direction.  However, as stated in Section 1, we do not intend to explain the nonlocality or the counterfactual non-definiteness of quantum mechanics by means of other concepts.  We interpret the equations derived from pure quantum mechanics.  The spin in either direction is not fixed in the initial state (\ref{eq:epr1}), and the spin in the specified direction of each electron is fixed in the state (\ref{eq:epr3}).  Transformation between equations (\ref{eq:epr1}) and (\ref{eq:epr3}) is due to the uncertainty relation.  However, we do not guess what occurs between them.  What we do attempt to illustrate in this study is how measurement processes can be understood within the scope of pure quantum mechanics.

\section{Double-slit experiment}
\label{sec:3}
\begin{figure}
\centering
\includegraphics[width=10cm]{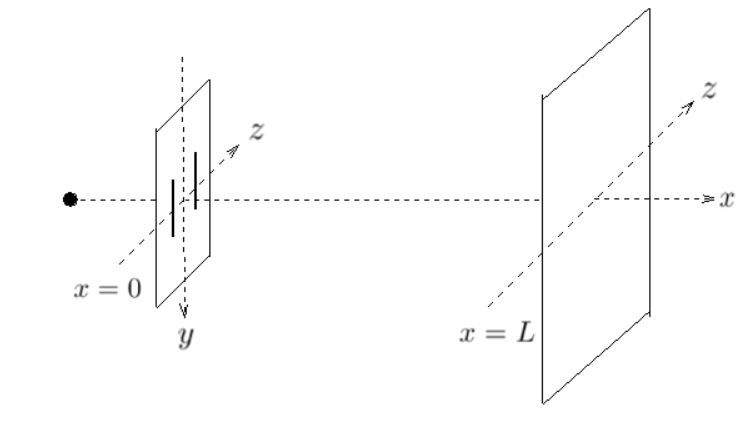}
\caption{Double-slit experiment}
\end{figure}
In this section, we examine the double-slit experiment, where electrons are emitted at certain intervals (Fig. 3).  The electrons travel along the $x$ axis and through a double slit in the $x=0$ plane, and finally they arrive somewhere on the screen in the $x=L$ plane.  Each slit is separated in the $y$ direction, and the two slits are positioned at $z=z_1$ and $z=z_2$.  We define the eigenstate $|z\rangle$ of the $z$ coordinate operator $\hat z$ as
\[
\hat z|z\rangle=z|z\rangle .
\]
Moreover, we define $|\psi\rangle$ as the state of the electron at $x=X\ (0\le X<L)$ and $|\psi_1\rangle$ and $|\psi_2\rangle$ as the states obtained via slits 1 and 2, respectively.  Then,
\begin{equation}
|\psi\rangle=|\psi_1\rangle + |\psi_2\rangle .
\end{equation}
The wave function $\psi (z)$ is defined as
\begin{equation}
|\psi\rangle=\int^{\infty}_{-\infty}dz\ \psi(z)|z\rangle,
\label{eq:joutai}
\end{equation}
with the normalization
\[
\int^{\infty}_{-\infty}dz\ |\psi(z)|^2=1.
\]
The probability density that the electron is observed at $x=X,\ z=z_0$ is
\begin{equation}
\begin{array}{rl}
\langle z_0\rangle&=\langle \psi|z_0\rangle\langle z_0|\psi\rangle\\
&=\int dz\int dz^{\prime}\ \psi^{\ast}(z^{\prime})\psi(z)\langle z^{\prime}|z_0\rangle\langle z_0|z\rangle\\
&=|\psi(z_0)|^2 .
\end{array}
\end{equation}
Then, $\psi_1(z)$ and $\psi_2(z)$ are defined in the same manner as (\ref{eq:joutai}) as 
\begin{equation}
|\psi_1\rangle=\int^{\infty}_{-\infty}dz\ \psi_1(z)|z\rangle,
\end{equation}
\begin{equation}
|\psi_2\rangle=\int^{\infty}_{-\infty}dz\ \psi_2(z)|z\rangle.
\end{equation}
Then,
\begin{equation}
\psi(z)=\psi_1(z)+\psi_2(z),\label{eq:hadoukannsuu}
\end{equation}
and the probability density $|\psi(z_0)|^2$ is written as
\begin{equation}
|\psi(z_0)|^2=|\psi_1(z_0)|^2+|\psi_2(z_0)|^2+2\Re \big(\psi_1(z_0)\psi_2^{\ast}(z_0)\big).\label{eq:kannshou}
\end{equation}
\subsection{Decoherence on the screen}
In this subsection, the electrons are not assumed to be observed at the slits.  In this case, the interference terms are included in the probability density $|\psi(z)|^2$, but each electron behaves as a particle on the screen. Therefore, we illustrate here that the density matrix of the electrons to be observed on the screen is not $|\psi\rangle\langle\psi|$ but is proportional to 
\[
\int dz|\psi(z)|^2|z\rangle\langle z|.
\]

In contrast to the what is stated in the previous sections, we must allow some uncertainty in the momentum in the $z$ direction, because we want to know the position in this direction.  Therefore, we define the averaged density matrix $\hat\rho_{\psi av}$ that describes the state of the electron to be observed on the screen as
\begin{equation}
\hat\rho_{\psi av}\equiv\frac{1}{2h\Delta p}\int^{\Delta p}_{-\Delta p}d\pi\hat\rho_{\psi}(\pi),
\end{equation}
where $\hat\rho_{\psi}(\pi)$ is defined as
\begin{equation}
\hat\rho_{\psi}(\pi)\equiv \hat \mathcal{T}(\pi)|\psi\rangle\langle \psi|\hat \mathcal{T}^{\dagger}(\pi),
\end{equation}
with 
\begin{equation}
\hat \mathcal{T}(\pi)=\exp\Big(\frac{i\hat z\pi}{\hbar}\Big).
\end{equation}
Here, $\hat \mathcal{T}(\pi)$ is the operator that changes the momentum and satisfies
\[
\langle p+\pi|\hat \mathcal{T}(\pi)|\psi\rangle =\langle p|\psi\rangle,
\]
where $|p\rangle$ is the eigenstate of the momentum in the $z$ direction.  Then,
\begin{equation}
\hat\rho_{\psi av}=\int^{\Delta p}_{-\Delta p}\frac{d\pi}{2h\Delta p}\int dz\int dz^{\prime}\psi^{\ast}(z^{\prime})\psi(z)\exp\Big(\frac{i(z-z^{\prime})\pi}{\hbar}\Big)|z\rangle\langle z^{\prime}.|\label{eq:ws}
\end{equation}

Here, because we want to know what is observed on the screen, we set
\[
\hbar\ll(z-z^{\prime})\Delta p, 
\]
which leads to
\begin{equation}
\int^{\infty}_{\Delta p}d\pi\exp\Big(\frac{i(z-z^{\prime})\pi}{\hbar}\Big)=0.
\end{equation}
By means of this equation, we can determine a delta-functional in (\ref{eq:ws}) as follows:
\begin{equation}
\begin{array}{rl}
\frac{1}{2\Delta p}\int\limits^{\Delta p}_{-\Delta p}d\pi\exp\Big(\frac{i(z-z^{\prime})\pi}{\hbar}\Big)&\sim\frac{1}{\delta(0)}\int\limits^{\infty}_{-\infty }d\pi\exp\Big(\frac{i(z-z^{\prime})\pi}{\hbar}\Big)\\
&=\frac{h}{\delta(0)}\delta (z-z^{\prime}),
\end{array}
\end{equation}
and (\ref{eq:ws}) becomes
\begin{equation}
\begin{array}{rl}
\hat\rho_{\psi av}&=\frac{1}{\delta (0)}\int dz\int dz^{\prime}\psi^{\ast}(z^{\prime})\psi(z)\delta (z-z^{\prime})|z\rangle\langle z^{\prime}|\\
&=\frac{1}{\delta (0)}\int dz|\psi(z)|^2|z\rangle\langle z|.\label{eq:ws2}
\end{array}
\end{equation}
Here, (\ref{eq:ws2}) is the desired form of the density matrix.  Because the state that (\ref{eq:ws2}) describes is a proper mixture, we can predict that each electron will behave as a particle.  However, because $|\psi (z)|^2$ is a probability density that includes the interference terms, the marks the electrons leave on the screen form a striped interference pattern.

\subsection{Decoherence near the slits}
Next, we examine the case in which the electrons are assumed to be observed near either of the slits, i.e., at $x=X\simeq 0$.  In this case, 
\begin{equation}
\psi_1(z)\psi_2^{\ast}(z)=0,
\end{equation}
from the definitions of $\psi_1(z)$ and $\psi_2(z)$.  Therefore, (\ref{eq:ws2}) becomes
\begin{equation}
\hat\rho_{\psi av}=\frac{1}{\delta (0)}\int dz\Big(|\psi_1(z)|^2+|\psi_2(z)|^2\Big)|z\rangle\langle z|,\label{eq:ws3}
\end{equation}
which shows that the interference terms have vanished.  If we observe the electron on the screen again, its probability density is not $|\psi(z)|^2$ but rather $|\psi_1(z)|^2+|\psi_2(z)|^2$.

\section{Conclusion}
\label{sec:4}
We have demonstrated in this study that decoherence is one of the inherent characteristics of pure quantum mechanics.  Therefore, we conclude that the quantum measurement process can be explained within the framework of pure quantum mechanics.  We believe that our study can be applied to a more general discussion on the quantum-to-classical transition.



\end{document}